%% file: index.tex
\documentclass[fleqn,10pt]{article}
\usepackage[margin=0.5in]{geometry}
\usepackage{braket}
\usepackage{mathtools}
\usepackage{hyperref}
\usepackage{graphicx, caption, subcaption}
\usepackage{float}
\usepackage[toc,page]{appendix}
\usepackage[T1]{fontenc}
\usepackage[utf8]{inputenc}
\usepackage{authblk}
\usepackage{amsmath}
\usepackage{color}
\usepackage{soul}

\newcommand{\raja}[1]{{\textcolor{blue}{#1}}} 

\newcommand{\norm}[1]{\left\lVert#1\right\rVert}
\newcommand{\comment}[1]{}

\title{Prime Factorization Using Quantum Variational Imaginary Time Evolution}

\author[1]{Raja Selvarajan} 
\author[1]{Vivek Dixit}
\author[2,3]{Xingshan Cui}
\author[3]{Travis S.~Humble\thanks{This manuscript has been authored by UT-Battelle, LLC, under contract DE-AC05-00OR22725 with the US Department of Energy (DOE). The US government retains and the publisher, by accepting the article for publication, acknowledges that the US government retains a nonexclusive, paid-up, irrevocable, worldwide license to publish or reproduce the published form of this manuscript, or allow others to do so, for US government purposes. DOE will provide public access to these results of federally sponsored research in accordance with the DOE Public Access Plan (http://energy.gov/downloads/doe-public-access-plan).}}
\author[1,3,*]{Sabre Kais}
\affil[1]{Department of Chemistry, Department of Physics and Astronomy, and Purdue Quantum Science and Engineering Institute, Purdue University, West Lafayette, IN 47907 USA}
\affil[2]{‘Department of Mathematics, Department of Physics and Astronomy, and Purdue Quantum Science and Engineering Institute, Purdue University, West Lafayette, IN 47907 USA’}
\affil[3]{Quantum Science Center Oak Ridge National Laboratory, Oak Ridge, TN}

\affil[*]{kais@purdue.edu}

\begin{document}

\maketitle

\input{sections/abstract}

\input{sections/introduction}

\input{sections/qite}

\input{sections/optimization}

\input{sections/example}

\input{sections/qfi}

\input{sections/simulation}

\input{sections/ibm}

\input{sections/conclusion}

\bibliographystyle{unsrt}
\bibliography{index}

\newpage

\begin{appendices}

    \input{sections/Appendix/mlp}

    \input{sections/Appendix/runtime}
\end{appendices}

\end{document}

%% file: sections/abstract.tex
{\fontfamily{ptm}\selectfont
\begin{abstract}
    The road to computing on quantum devices has been accelerated by the promises that come from using Shor's algorithm to reduce the complexity of prime factorization. However, this promise hast not yet been realized due to noisy qubits and lack of robust error correction schemes. Here we explore a promising, alternative method for prime factorization that uses well-established techniques from variational imaginary time evolution. We create a Hamiltonian whose ground state encodes the solution to the problem and use variational techniques to evolve a state iteratively towards these prime factors. We show that the number of circuits evaluated in each iteration scales as $O(n^{5}d)$, where $n$ is the bit-length of the number to be factorized and $d$ is the depth of the circuit. We use a single layer of entangling gates to factorize several numbers represented using 7, 8, and 9-qubit Hamiltonians. We also verify the method's performance by implementing it on the IBMQ Lima hardware.
\end{abstract}}

%% file: sections/introduction.tex
\section{Introduction}

Quantum computation is likely to revolutionize how computation is performed in the field of science, engineering, and finance. Computations are performed on quantum states that make use of superposition and entanglement to allow for speedups. Future potential applications include cryptography \cite{Bernstein2009}, search problems \cite{Andris}, simulation of quantum systems \cite{Preskill_2018}, quantum annealing \cite{dixit2020training}, machine learning \cite{ieee_vivek}, computation biology \cite{comp_bio}, quantum materials \cite{suresh2021}, and problems in optimization \cite{cao2016solving}. Refer \cite{kais2014quantum} for an extensive introduction into the field of quantum computation. Major efforts toward scaling the current algorithm focuses on developing error correction and mitigation schemes, as well as designing operations that make use of fewer ancilla qubits and gate operations \cite{brown2015reducing}. In line with the current major developments, we investigate a more practical near-term scheme for prime factorization that is likely to achieve good results on noisy qubits.

Prime factorization involves expressing a composite number as the product of its prime factors. For generic large numbers that lack any structure, the quadratic sieve is the most commonly used technique. This can be computationally expensive and is exploited in RSA cryptography to guarantee information security over networks. From Shor's~\cite{Shor_1997} work on period finding, it was shown that one could exploit the quantum Fourier transform to compute factors in steps that scaled polynomial in the number of bits. Subsequently, Vandersypen et al.~\cite{Vandersypen_2001} realized this experimentally by factorizing $N=15$ using spin-$1/2$ nuclei as qubits and then Lucero et al.~\cite{Lucero_2012} by using superconducting qubits. A simplified version of  Shor's algorithm was worked out for products of Fermat primes (3, 5, 17, 257 and 65537) by Geller et al.~\cite{geller2013factoring}. Jian et al.~\cite{jiang2018quantum} proposed an alternative method to compute factors by solving an optimization problem using quantum annealing demonstrated on the D-Wave quantum annealer. The largest experimental realization of general method factorization schemes includes Shor's algorithm to factorize 21~\cite{Lucero_2012} and optimization using the D-Wave quantum annealer to factorize 223357~\cite{jiang2018quantum}. In addition, large numbers with specific properties have also been factorized by exploiting structure contained in the number. In Ref.~\cite{dattani2014quantum}, the authors use a multiplication table to factorize 56153 using only 4 qubits.
Despite being able to factorize large numbers with relatively few qubits by exploiting the structure of the number, these methods do not reveal the power of quantum computers against generic numbers. Studies have not been made with respect to convergence on the solution with increasing the number of qubits. 
\par 
In this paper we explore how one could use imaginary time evolution to factorize numbers with relatively higher probability. Imaginary time evolution has long been used as a theoretical tool in physics to compute ground-state wave-functions~\cite{magnus}. Shingu et al.~\cite{shingu2020boltzmann} show how using imaginary time evolution one could train a Boltzmann network efficiently, computing the exact model independent term in the training. Recently, Motta et al.~\cite{Motta_2019} exploits the Quantum Imaginary Time Evolution (QITE) algorithm to determine eigenstates and thermal states on a quantum computer. McArdle et al.~\cite{McArdle_2019} showed how one could make use of variational circuits to create states that represent the dynamics of the imaginary time evolution. They use it to compute the ground state wavefunction of hydrogen and lithium hydride, while Yeter-Aydeniz et al.~\cite{2021yeterQITE} demonstrate the QITE algorithm as a quantum computing benchmark for computational chemistry methods. 
\par
Herein, we develop an optimization function using the method referenced at~\cite{burges2002factoring} and then use it as a Hamiltonian to perform imaginary time evolution on a uniform superposition of all possible considered solutions to the factorization problem. We employ variational circuits to prepare a quantum state that encodes the solution and classically train all parameters. Simulations using Python Numpy packages and IBM-QASM are used to verify the performance. \comment{We compare it against Variational Quantum Eigensolver (VQE) to show why QITE performs better and avoids the issue of barren plateau~\cite{McClean_2018} in exploring the landscape.} The robustness of the techniques against noise is verified on IBMQx2 hardware for up to 5 qubits allowing factorization of numbers up to 91 with a probability over $73\%$. To the best of our knowledge this is the largest number that has been factorized on a quantum circuit using a general purpose algorithm that does not exploit any structural properties of the number. Each iteration involves evaluating a number of circuits that scales as $O(n^5d)$ where $n$ is the bit-string length of the number to be factorized and $d$ is the circuit depth. A few reasons make factorization an ideal candidate to be solved using Imaginary time evolution with variational circuits. The number of terms in the Hamiltonian expansion scales polynomial with respect to the number of qubits. The amplitude of coefficients in the state vector is real, which simplifies the parameterization of the landscape to be explored and updates to be made. Each iteration only needs to amplify the amplitude of the solution rather than exactly simulate the imaginary time evolution. These factors make factorization using the QITE algorithm a good candidate to demonstrate the practicality of quantum circuits in near term devices in the NISQ~\cite{Preskill_2018} era.

%% file: sections/qite.tex
\section{Quantum Imaginary Time Evolution}
The Schr\"odinger equation for a closed system describes dynamics according to some Hamiltonian that governs it. Allowing for time to be complex, we are able to create thermal states of specific temperature starting from a maximally mixed configuration, i.e, $\rho_{T=1/\tau} = e^{-H\tau}/$Tr$[e^{-H\tau}]$. Preparing a system at low temperature or alternatively letting the system evolve to large imaginary time values we can more often sample the ground state configuration for a given Hamiltonian. We shall make use of this property to encode the required solution of the problem we intend to solve in the later sections.

The quantum state we intend to prepare using QITE is given by,
\begin{equation}\label{eq:def}
    \ket{\psi(\tau)} = N(\tau) e^{-H\tau}\ket{\psi(0)}
\end{equation}
where $N(\tau)= 1/\sqrt{\bra{\psi(0)}e^{-2H\tau}\ket{\psi(0)}}$ is a normalization factor. Equation \ref{eq:def} satisfies the Wick rotated Schrodinger equation
\begin{equation} \label{eq:evol}
    \frac{\partial\ket{\psi(\tau)}}{\partial\tau} = - (H -E_\tau) \ket{\psi(\tau)}
\end{equation}
where $E_{\tau}=\bra{\psi(\tau)}H\ket{\psi(\tau)}$.

Let $\ket{\phi(\tau)}$ be a state that is prepared by applying a series of unitary gates as follows
\begin{equation}\label{eq:ansatz}
    \ket{\phi(\tau)} = U_N(\theta_N)U_{N-1}(\theta_{N-1})U_{N-2}(\theta_{N-2})U_{N-3}(\theta_{N-3})...U_1(\theta_1)\ket{0}
\end{equation}

We choose the initial parameters to create the state on which the evolution is to be performed. We demand that equation \eqref{eq:ansatz} is an approximate solution to equation \eqref{eq:evol} by demanding the norm of the variation to vanish, i.e,
\begin{equation}\label{eq:norm} 
    \delta \norm{\left(\frac{\partial}{\partial\tau} + H - E_\tau\right) \ket{\phi(\tau)}} =0
\end{equation}

Solving the above equation using McLachlan's variational principle (details expanded in the Appendix) we obtain
\begin{equation}\label{eq:grad}
    \sum A_{ij} \dot{\theta_j} = C_i
\end{equation}
where
\begin{equation}\begin{split}
     A_{ij}= \frac{\partial\bra{\phi(\tau)}}{\partial\theta_i}{\frac{\partial\ket{\phi(\tau)}}{\partial\theta_j}} \\
     C_i = -  \frac{\partial\bra{\phi(\tau)}}{\partial\theta_i}H\ket{\phi(\tau)}
\end{split}
\end{equation}

The derivatives on the state vectors with respect to the parameters of the circuit can be simplified into a sum of other basic circuit functions. For instance, the derivative of the $R_x$, $R_y $ and $R_z$ rotation gates with respect to their parameter amounts to a mere shift of the parameter by an angle of $\pi$ radians. We can further express $H$ as a sum of a string of tensor products of Pauli operators allowing both $A$ and $C$ to be computed as the sum of terms that take the form $a \Re(e^{i\phi}\bra{0}U\ket{0})$, where $a$ is some real number and $\phi$ is a real phase. Terms of this form can be efficiently computed on quantum hardware using the Hadamard test.

Evaluating the gradients using equation \eqref{eq:grad}, we can then proceed to use gradient descent to update the parameters of the circuit as follows
\begin{equation}\label{eq:theta_upd}
    \vec{\theta}(\tau +\delta\tau) = \vec{\theta}(\tau)  + \dot{\vec{\theta}}(\tau)\delta\tau
\end{equation}
where $\dot{\vec{\theta}}(\tau) =  A^{-1}(\tau)C(\tau)$. Using sufficient layers to the variational circuit and small $\delta\tau$ time iterations for updating our parameters, we can prepare states that closely emulate the imaginary time evolution on the state $\ket{\psi(0)}$. We would like to note that the circuit prepares the final state from a fixed starting state rather than from the state at $\delta t$ time step before evolution as this would correspond to a non-unitary transformation on the Hilbert space of qubits.

%% file: sections/optimization.tex
\section{Prime Factorization to Binary Optimization}

\quad The problem of prime factorization involves expressing a number as a product of its primes. Here we shall focus specifically on biprimes, product of $2$ prime numbers. Shor's algorithm aims to solve this problem by reducing it to a period finding problem of the function $a^x$ mod $N$, where $a$ is any random number and $N$ is the number to be factorized. This algorithm uses a quantum subroutine that exploits the power of quantum Fourier transform as a part of a modified phase estimation algorithm. The largest number factorized using Shor's algorithm is 21. Being highly sensitive to discrete errors, Shor's algorithm fails to scale without viable error correction schemes~\cite{devitt2005investigating}.

Let the number $N$ be given by a product of $2$ prime numbers $p$ and $q$. We model the problem of finding $p$ and $q$ given $N$ as an optimization problem cast as the following cost function
\begin{equation}\label{eq:opt}
    H(p,q) = (N-p \times q)^2
\end{equation}

Note that in equation \ref{eq:opt}, $H(p,q)$ has global minima when $p$,$q$ factorize $N$ with a minimum value of 0. The cost function can thus be treated as a Hamiltonian whose ground state encodes the solution to our problem. We express both $p$ and $q$ as a binary string summation which will allow us to transfer to a qubit representation space. Assuming $p$ has an $m$ bit  representation and $q$ has an \raja{$l$} bit representation, we get
\begin{equation}\label{eq:binexp}
    \begin{aligned}
        &p = 2^{m-1}p_{m-1} +  2^{m-2}p_{m-2} \ldots 2^{0}q_{0} \\
        &q = 2^{l-1}q_{l-1} +  2^{l-2}q_{l-2} \ldots 2^{0}q_{0} 
    \end{aligned}
\end{equation}

Note that equation \ref{eq:opt} has a trivial solution $N=N \times 1$. To exclude this solution, we can restrict the search space of our solution by using a lower bound. The smallest prime factor we are looking for is greater than $2$ and less than $N/2$. Thus using $N-1$ qubits to represent $p$ and $q$ we can discard the trivial solution from showing up in our simulation. Since both $p$ and $q$ are prime numbers, we are allowed to set the bit $0$ to be 1 in either expression making it indivisible by 2. To move to a scaled spin representation we transform all the binary variables $b_i$ that represents either the $p_i$'s or $q_i$'s to $ b_i =  (s_i + 1)/2$. In this representation, $s_i$ takes the value $\pm 1$ such that $s_i=1$ maps to $b_i=1$ and $s_i=-1$ maps to $b_i=0$. We thus convert a factorization problem into an optimization problem over the spin variables.\\

Minimizing this optimization function shall result in the solution we seek. We use imaginary time evolution to evolve a uniform superposition over all possible solutions. The circuit parameters are updated according to equation \eqref{eq:grad} and \eqref{eq:theta_upd}. With every iteration that is performed, the output state evolves through a time step of $\delta\tau$. After sufficiently large enough time $T$, we expect the ground state with the least cost to prepare. For $\norm{H}_{1}T>> 1$ we get,
\begin{equation}
    e^{-HT} \ket{++.. +} \approx \ket{\tilde{p}}\ket{\tilde{q}}
\end{equation}
where $\tilde{p}$ and $\tilde{q}$ represent the bit string solution of the binary representation of the numbers that multiply to give $N$. The evolution is achieved by using a variational circuit ansatz to prepares states that mimic the evolution of the system from a given starting state (as described in the previous section). We show in the Appendix that the computational cost in performing one iteration of QITE is $O(n^5d)$, where $n$ is the bit-length of the number to be factored and $d$ is the depth of the circuit.

%% file: sections/example.tex
\newpage
\section{Demonstrating an Example}
\begin{flushleft}
    Here we show how to factorize $N = 15$ as
    \begin{equation}
        15 =  5 \times 3
    \end{equation}

    Expressing the factors 5 and 3 as binary variables and setting $p_0 =1$, $q_0 = 1$ yields
    \begin{equation}
        \begin{aligned}
            &5 = 4x_1 + 2x_0 + 1 \\
            &3 = 2x_2 + 1 
        \end{aligned}
    \end{equation}

    with the correct results defined by $x_0=0, x_1=1, x_2=1$. The  corresponding cost function $H(p,q)$ is cast as 
    \begin{equation}
        H(x_0,x_1,x_2) = [15 - (4x_1 + 2x_0 + 1)(2x_2 + 1)]^2 
    \end{equation}
    
    Expanding the cost function and making use of $x_i^2 =x_i$ yields
    \begin{equation}
    H(x_0,x_1,x_2) = 196 - 52 x_2 - 52 x_0 - 56 x_2 x_0 - 96 x_1 - 48 x_2 x_1 + 16 x_0 x_1 + 128 x_0 x_1 x_2
    \end{equation}
    
    Mapping the binary variables to spin variables yields
    \begin{equation}
    H(s_0,s_1,s_2) = 90 - 36 s_2 - 40 s_1 - 20 s_0 + 2 s_2 s_0 + 4 s_2 s_0+ 20 s_0 s_1 + 16 s_0 s_1 s_2
    \end{equation}

The classical binary variables $s_i$ in the Hamiltonian is replaced by the Pauli-z spin operator to get a quantum Hamiltonian Operator that is then acted upon the full superposition state.
 Using a time step $\delta \tau=0.1$ , we make $10$ iterations of QITE to reach $T=1$. With a probability of greater than $90\%$, we obtain upon measurement
    \begin{equation}\label{eq:sol}
        e^{-HT} \ket{+++} \approx \ket{110}
    \end{equation}
    where the above expression is written in the computational basis. The output equation \eqref{eq:sol} when mapped back to the binary variables gives
    \begin{equation}
        x_0 = 0,     x_1 = 1,     x_2=1
    \end{equation}
    
    We thus obtain the factors of 3 and 5 as expected. In our simulations, we shall make use of no more than the number of qubits needed to represent our solution in the optimization. This is to limit the numbers of qubits used, maximize the computational efficiency and reduce the accumulated errors.
    
    \end{flushleft}

%% file: sections/qfi.tex
\section{Reformulating using QFI and Gradients}

\begin{figure}[h]
    \centering
    \includegraphics[height=0.30\textwidth]{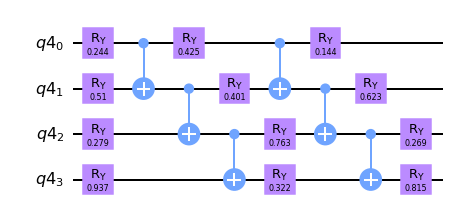}
    \caption{Variational circuit used to prepare the quantum state for imaginary time evolution. The parameters have been randomly initialized}
    
    \label{fig:vqe_circref}
\end{figure}

We generate the output quantum state prepared using QITE with the ansatz shown in figure~\ref{fig:vqe_circref}. Notice that the circuit consists of $R_Y$ rotation gates applied to each qubit followed by a layer of $CNOT$ gates that help with the entanglement of the qubits. Using only $R_Y$ gates ensures that we maintain real amplitudes for the state when expressed in the computation basis. This helps with closely following the dynamics of QITE without introducing any nontrivial phase values.\\ 

We show how for such an ansatz, equation \ref{eq:grad} can be re-expressed in terms of the quantum Fisher information (QFI) and the gradient of Hamiltonian expectation with respect to the current state. In classical probability, the Fisher information characterizes how much a probability distribution varies by changing a parameter that characterizes a distribution. This can be generalized and extended to talk about quantum states where one characterizes how much a state changes with respect to the parameter that governs it. Refer to \cite{PETZ_2011} for a brief introduction to quantum Fisher information. Given a quantum state $\ket{\psi(\theta_1,\theta_2 ... \theta_n)}$, where $\theta_i$ represents the governing parameters, the QFI is given by
\begin{equation}
    F_{ij} = 4 \Re(\braket{\psi_i|\psi_j} -\braket{\psi_i|\psi}\braket{\psi|\psi_j})
\end{equation}
where $\ket{\psi_j}=\frac{d\ket{\psi(\theta_j)}}{d\theta}$ . For the ansatz being used to generate the state in QITE, we note that $\braket{\psi_i|\psi} = \Re(\braket{\psi_i|\psi})$. This results in the second term vanishing due to the constant normalization of the state i.e, $\braket{\psi|\psi}=1$. Hence we get $F_{ij} = 4M_{ij}$ where $M_{ij} =\Re(\braket{\psi_i(\theta(\tau))|\psi_j(\theta(\tau))})$, cf.~Eq.~(26) in  Appendix A.

Furthermore, since the Hamiltonian is an Hermitian operator, we can express the right hand side of equation \eqref{eq:grad} as follows
\begin{equation}
    C_i = \Re(\bra{\psi}H\ket{\psi_i}) = \frac{d}{d\theta_i}(\bra{\psi}H\ket{\psi})/2
\end{equation}
We thus obtain $\sum\limits_{j} F_{ij} \Dot{\theta_j} = - 2\frac{d}{d\theta_i}\braket{H}_\psi$. Both the quantum Fisher information of a given circuit and the gradient of an observable with respect to a state parameter can be directly computed in the IBM Qiskit Aqua framework.

%% file: sections/simulation.tex
\section{Results on Simulation}

\begin{figure}[H]
    \centering
    \begin{subfigure}[b]{0.48\textwidth}
        \includegraphics[width=\linewidth]{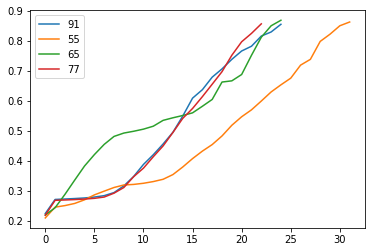}
        \caption{Amplitude of correct solution for various 5-qubit examples.}
        \label{fig:5qubit_simulations}
    \end{subfigure}
   \begin{subfigure}[b]{0.48\textwidth}
       \includegraphics[width=\linewidth]{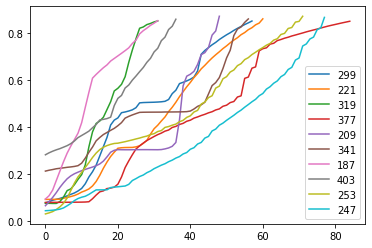}
       \caption{Amplitude of correct solution for various 7-qubit examples.}
       \label{fig:7qubit_simulations}
   \end{subfigure}

   \begin{subfigure}[b]{0.48\textwidth}
    \includegraphics[width=\linewidth]{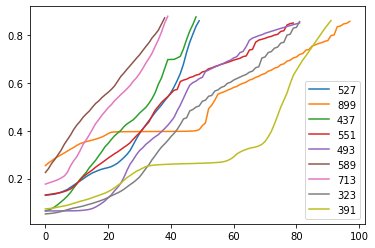}
    \caption{Amplitude of correct solution for various 8-qubit examples.}
    \label{fig:8qubit_simulations}
    \end{subfigure}
   \begin{subfigure}[b]{0.48\textwidth}
       \includegraphics[width=\linewidth]{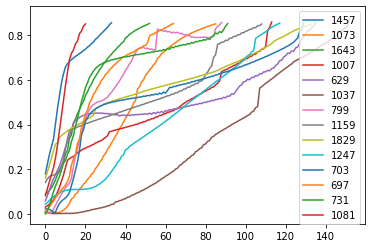}
       \caption{Amplitude of correct solution for  various 9-qubit examples.}
       \label{fig:9qubit_simulations}
   \end{subfigure}

   \caption{Numerical simulations of QITE for 7-, 8-, and 9-qubit factorization examples. The vertical axis indicates the amplitude of the solution in the computational basis and the horizontal axis indicates the number of iterations made. The curves in each subplot show how the amplitude of the solution corresponding to various numbers increases with iterations. The amplitude threshold has been set to 0.85.}
   \label{fig:fact_simulations}
\end{figure}

Smaller numbers were factorized on Qiskit using the QASM simulator. For 8 qubits and more, the simulations were run using the Numpy package. The amplitude threshold for the correct solution has been set to $0.85$, this amounts to a probability of $73\%$ to get the right solution when measured in the computational basis. We have restricted our ansatz to a single layer in addition to the base layer, in light of exploring shallow circuits for the near term quantum computers. The initial circuit parameters have been randomly sampled to help avoid valleys during the training and allow for faster convergence with shallow circuits employed. We have noticed that it also helps in suppressing one of the solutions in the case of symmetrically equivalent solutions (for instance, 391 can be factorized as $17 \times 23$ and $23 \times 17$), allowing one solution to quickly reach the threshold.Figure \ref{fig:fact_simulations} plots the amplitude of the solution in the computational basis against the number of iterations for several values of $N$ that are represented using 5, 7, 8, and 9-qubit Hamiltonians. Note the amplitude of the solution to the factorization keeps raising with increasing the number of iterations. The figure also shows that the average number of iterations required to factorize increases with the number of qubits though we have not been able to analytically bound it. A few instances involve the amplitude of the solution flattening before continuing to raise any further, making it significantly slow. In table \ref{tab:iteration_list}, we note that despite some numbers being close to each other (for instance 1007 and 1081), the number of iterations for convergence to the solution seem to be widely varying. This can be attributed to the differences in bit string representation and thus the created cost function, and also to the fact that the parameters are being randomly initialized. Results on actual hardware for smaller instances have been presented in section 8. 

\begin{table}
    \begin{center}
    \begin{tabular}{||c c||} 
    \hline
    Number factorized &  Number of iterations \\ [0.5ex] 
    \hline\hline
    55 & 31  \\ 
    \hline
    65 & 23  \\
    \hline
    77 & 21  \\
    \hline
    91 & 23  \\
    \hline
    187 & 30 \\
    \hline
    209 & 48  \\ 
    \hline
    221 & 61 \\
    \hline
    247 & 77  \\
    \hline
    253 & 72  \\
    \hline
    299 & 58 \\
    \hline
    319 & 30 \\ 
    \hline
    341 & 57 \\ 
    \hline
    377 &  84 \\
    \hline
    403 & 36 \\
    \hline
    323 & 81  \\
    \hline
    391 & 89 \\
    \hline
    437 & 48  \\ 
    \hline
    493 & 80 \\
    \hline

   \end{tabular}
\quad
\begin{tabular}{||c c||} 
    \hline
    Number factorized &  Number of iterations \\ [0.5ex]
    \hline\hline
    527 & 47  \\
    \hline
    551 & 77  \\
    \hline
    589 & 38  \\
    \hline
    629 & 132  \\
    \hline
    697 & 59  \\
    \hline
    703 & 31  \\
    \hline
    713 & 39 \\
    \hline
    731 & 56  \\
    \hline
    799 & 86  \\
    \hline
    899 & 96 \\
    \hline
    1007 & 21  \\
    \hline
    1037 & 155  \\
    \hline
    1081 & 112  \\
    \hline
    1159 & 103  \\
    \hline
    1247 & 112  \\
    \hline
    1457 & 128 \\
    \hline
    1643 & 88  \\
    \hline
    1829 & 130  \\
    \hline
\end{tabular}
\end{center}
\caption{\label{tab:iteration_list} Table lists the numbers that were factorized alongside the number of iterations for convergence}
\end{table}

%% file: sections/ibm.tex
\section{Results on IBM Hardware}
\begin{figure}[H]
    \centering
    \includegraphics[width=0.5\textwidth]{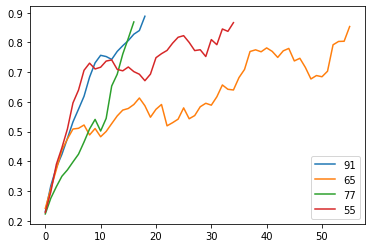}
    \caption{5-qubit factorization on IBMQ-lima. The vertical axis indicates the amplitude of the solution in the computational basis and the horizontal axis indicates the number of iterations made. The curves in each subplot, show how the amplitude of the solution corresponding to various numbers increase with iterations. The amplitude threshold has been set to 0.85}
    \label{fig:IBMQ}
\end{figure}

We have made use of the publicly available IBM hardware to simulate factorization for up to 5 qubits. We start the output state with a uniform superposition over all possible solutions. This is straightforwardly instantiated by setting the parameters of $R_y$ gates in the base layer to be $\frac{\pi}{2}$ and the remaining parameters to be all $0$. We have avoided invoking the Hadamard test to compute the $M$ matrix and $C$ vector as this resulted in significant error accumulation. Instead, we have reformulated the computation as indicated in section 5 and have made use of built-in Qiskit modules. The Gradient module with parameter shift method has been used to compute the $C$ vector, while the QFI module with overlapping block diagonal method has been used to compute the $M$ matrix. Please refer to the Qiskit source documentation \cite{Qiskit} for details on the implementation of these modules. Figure \ref{fig:IBMQ} shows the results of factorizing the numbers $N = $ 55, 65, 77 and 91 on IBMQ-lima hardware that supports 5 qubits. Unlike the simulations that hardly show any oscillation in the amplitude, we see oscillations being introduced due to noise when run on the hardware. The convergence of the solution in the presence of hardware noise makes this method a suitable candidate for solving factorization and other similarly framed optimization problems in the near term quantum computers.

%% file: sections/conclusion.tex
\section{Conclusions}
We have shown how imaginary time evolution can be used to perform optimization and we have demonstrated this for the case of prime factorization. We have shown numerical adn experimental results from the factorization of several numbers represented by 7, 8, and 9-qubit Hamiltonians. We have shown that imaginary time evolution significantly populates the probability of measuring the correct solution to the factorization problem when run sufficiently long. \comment{A comparison study against the most popular VQE technique shows that QITE performs significantly better even in the regime of shallow networks and avoids getting stuck exploring barren plateaus.} The observed performance of the method on the IBM-lima hardware reveals a robustness towards noise and works as a proof of concept in the case of a limited number of qubits. Methods on how to bound the number of iterations analytically can broaden the scope for this method on near-term quantum computers for very large numbers.

\section{Acknowledgement}
We would like to thank Dr. Mario Motta and Dr. Stefan Woerner of IBM for useful discussion. We would like to acknowledge the support from National Science Foundation under award number 1955907. This material is based upon work supported by the U.S. Department of Energy, Office of Science, National Quantum Information Science Research Centers, Quantum Science Center.

\section{Author contributions statement}
S.K., T.H. and X.C. designed the research problem, R.S. performed hardware testing and wrote the paper, R.S and V.D coded for simulations using Python Numpy. All authors helped in revising the paper draft.

%% file: sections/Appendix/mlp.tex
\section{McLachlan's Principle}
    We are borrowing the derivation from \cite{Yuan_2019} and including it here for completion.

\begin{equation}
    \delta \norm{(\frac{d}{d\tau}+ A )\ket{\psi(\tau)} } = 0 
\end{equation}

\begin{flushleft}
    Here we take $A$ to be Hermitian, which is true of the Hamiltonian, and work with the variation of the norm square
\end{flushleft}

\begin{flushleft}
\begin{equation}
    \norm{(\frac{d}{d\tau}+ A )\ket{\psi(\theta(\tau))} }^2 =((\frac{d}{d\tau}+ A )\ket{\psi(\theta(\tau)))})^{*}((\frac{d}{d\tau}+ A )\ket{\psi(\theta(\tau)))})
\end{equation}

Then, taking $\theta$ to be real, this yields

\begin{equation}
    \braket{\psi_i(\theta(\tau))|\psi_j(\theta(\tau))}\dot{\theta_i}\dot{\theta_j}+ \bra{\psi_i(\theta(\tau)))} A \ket{\psi(\theta(\tau)))}\dot{\theta_i} + \bra{\psi(\theta(\tau)))} A \ket{\psi_i(\theta(\tau)}\dot{\theta_i} + \bra{\psi(\theta(\tau)))}A^{2}\ket{\psi(\theta(\tau)))}
\end{equation}
where we have used the notation $\ket{\psi_i(\theta)}= \frac{d\ket{\psi(\theta)}}{d\theta_i}$. Computing the variation for a small time interval, we get
\begin{equation}
    (\braket{\psi_i(\theta(\tau))|\psi_j(\theta(\tau))}\dot{\theta_j}+ \bra{\psi_j(\theta(\tau)))} \frac{d\ket{\psi(\theta(\tau)}}{d\theta_i}\dot{\theta_j}  +\bra{\psi(\theta(\tau)))} A \ket{\psi_i(\theta(\tau)} + \bra{\psi_i(\theta(\tau)))} A \ket{\psi(\theta(\tau)))} )\delta\dot{\theta_i}
\end{equation}
Then, setting the variation to vanish for independent $\theta_i$, equates to the coefficients vanishing and yields
\begin{equation}
\Re(\braket{\psi_i(\theta(\tau))|\psi_j(\theta(\tau))})\dot{\theta_j} = -\Re(\bra{\psi(\theta(\tau)))} A \ket{\psi_i(\theta(\tau)})
\end{equation}
We next substitute $A=H-\bra{\psi(\theta)}H\ket{\psi(\theta)}$ and use the fixed normalization of the state with that $\Re(\bra{\psi(\theta)}\ket{\psi_i(\theta)})=0 $ to yield
\begin{equation}
\Re(\braket{\psi_i(\theta(\tau))|\psi_j(\theta(\tau))})\dot{\theta_j} = -\Re(\bra{\psi(\theta))} H \ket{\psi_i(\theta(\tau)})
\end{equation}
Finally, we get $M_{ij}\dot{\theta}=C_{i}$ with $M_{ij} =\Re(\braket{\psi_i(\theta(\tau))|\psi_j(\theta(\tau))}) $ and $C_{i}= \Re(\bra{\psi(\theta))} H \ket{\psi_i(\theta(\tau)})$

\end{flushleft}

%% file: sections/Appendix/runtime.tex
\section{Computations per iteration of the QITE}
Let the number $N$ to factorized be represented by $n$ bits. Each of the factors will be upper bounded by an $n$ bit representation. The Hamiltonian thus takes the following form
\begin{equation}
     H = [N-(2^{n-1}p_{n-1} + ..+ 2^{0}p_{0})(2^{n-1}q_{n-1} + ..+ 2^{0}q_{0})]^2
\end{equation}
Expanding this Hamiltonian, we obtain a constant term, $n^2$ terms of the type $p_iq_j$, $n^2(n-1)/2$ terms of the form $p_iq_jq_k$, $n^2(n-1)/2$ terms of the form $p_ip_jq_k$ and $n^2(n-1)^2/4$ terms of the form $p_ip_jq_kq_l$. The constructed Hamiltonian thus contains $ [n(n+1)/2]^2 +1 \approx O(n^4)$ terms in it. Note that the cost of computing the coefficients is also of the same order, given the number of terms is of $O(n^4)$, if one were to ignore any simplifications that are to be carried.

When we restrict to a $d$-layer Ry circuit, we update $nd$ circuit parameters with every algorithm iteration. This involves computing an $M$ matrix of size $nd$ and a C vector of size $nd$. Computing each entry in a C vector involves working with each term of the Hamiltonian independently and summing them up to the end, thus we have $O(n^5d)$ circuits to evaluate for C and $O(n^2d^2)$ circuits to evaluate for M. Taking the circuit depth to grow slower than $n^3$, we get a total of $O(n^5d)$ circuits to compute every iteration. To keep the overall run time of the QITE algorithm as polynomial, it is important to achieve convergence to the correct solution in polynomial number of steps.

%% file: index.bbl
\begin{thebibliography}{10}

\bibitem{Bernstein2009}
Daniel~J. Bernstein.
\newblock {\em Introduction to post-quantum cryptography}, pages 1--14.
\newblock Springer Berlin Heidelberg, Berlin, Heidelberg, 2009.

\bibitem{Andris}
Andris Ambainis.
\newblock Quantum search algorithms.
\newblock {\em ACM SIGACT News}, 35:22--35, 05 2005.

\bibitem{Preskill_2018}
John Preskill.
\newblock Quantum computing in the {NISQ} era and beyond.
\newblock {\em Quantum}, 2:79, Aug 2018.

\bibitem{dixit2020training}
Vivek Dixit, Raja Selvarajan, Muhammad~A. Alam, Travis~S. Humble, and Sabre
  Kais.
\newblock Training and classification using a restricted {B}oltzmann machine on
  the {D-W}ave 2000{Q}.
\newblock {\em preprint arxiv:2005.03247}, 2020.

\bibitem{ieee_vivek}
Vivek Dixit, Raja Selvarajan, Tamer Aldwairi, Yaroslav Koshka, Mark~A. Novotny,
  Travis~S. Humble, Muhammad~A. Alam, and Sabre Kais.
\newblock Training a quantum annealing based restricted boltzmann machine on
  cybersecurity data.
\newblock {\em IEEE Transactions on Emerging Topics in Computational
  Intelligence}, pages 1--12, 2021.

\bibitem{comp_bio}
Carlos Outeiral, Martin Strahm, Jiye Shi, Garrett~M. Morris, Simon~C. Benjamin,
  and Charlotte~M. Deane.
\newblock The prospects of quantum computing in computational molecular
  biology.
\newblock {\em WIREs Computational Molecular Science}, 11(1):e1481, 2021.

\bibitem{suresh2021}
Shree~Hari Sureshbabu, Manas Sajjan, Sangchul Oh, and Sabre Kais.
\newblock Implementation of quantum machine learning for electronic structure
  calculations of periodic systems on quantum computing devices.
\newblock {\em preprint arXiv:2103.02037}, 2021.

\bibitem{cao2016solving}
Yudong Cao, Shuxian Jiang, Debbie Perouli, and Sabre Kais.
\newblock Solving set cover with pairs problem using quantum annealing.
\newblock {\em Scientific reports}, 6(1):1--15, 2016.

\bibitem{kais2014quantum}
S.~Kais, K.B. Whaley, A.R. Dinner, and S.A. Rice.
\newblock {\em Quantum Information and Computation for Chemistry}.
\newblock Advances in Chemical Physics. Wiley, 2014.

\bibitem{brown2015reducing}
Katherine~L Brown, Anmer Daskin, Sabre Kais, and Jonathan~P Dowling.
\newblock Reducing the number of ancilla qubits and the gate count required for
  creating large controlled operations.
\newblock {\em Quantum Information Processing}, 14(3):891--899, 2015.

\bibitem{Shor_1997}
Peter~W. Shor.
\newblock Polynomial-time algorithms for prime factorization and discrete
  logarithms on a quantum computer.
\newblock {\em SIAM Journal on Computing}, 26(5):1484–1509, Oct 1997.

\bibitem{Vandersypen_2001}
Lieven M.~K. Vandersypen, Matthias Steffen, Gregory Breyta, Costantino~S.
  Yannoni, Mark~H. Sherwood, and Isaac~L. Chuang.
\newblock Experimental realization of shor’s quantum factoring algorithm
  using nuclear magnetic resonance.
\newblock {\em Nature}, 414(6866):883–887, Dec 2001.

\bibitem{Lucero_2012}
Erik Lucero, R.~Barends, Y.~Chen, J.~Kelly, M.~Mariantoni, A.~Megrant,
  P.~O’Malley, D.~Sank, A.~Vainsencher, J.~Wenner, and et~al.
\newblock Computing prime factors with a josephson phase qubit quantum
  processor.
\newblock {\em Nature Physics}, 8(10):719–723, Aug 2012.

\bibitem{geller2013factoring}
Michael~R. Geller and Zhongyuan Zhou.
\newblock Factoring 51 and 85 with 8 qubits.
\newblock {\em preprint arxiv:1304.0128}, 2013.

\bibitem{jiang2018quantum}
Shuxian Jiang, Keith~A. Britt, Alexander~J. McCaskey, Travis~S. Humble, and
  Sabre Kais.
\newblock Quantum annealing for prime factorization.
\newblock {\em Scientific Reports}, 8:17667, 2018.

\bibitem{dattani2014quantum}
Nikesh~S. Dattani and Nathaniel Bryans.
\newblock Quantum factorization of 56153 with only 4 qubits, 2014.

\bibitem{magnus}
Wilhelm Magnus.
\newblock On the exponential solution of differential equations for a linear
  operator.
\newblock {\em Communications on Pure and Applied Mathematics}, 7(4):649--673,
  1954.

\bibitem{shingu2020boltzmann}
Yuta Shingu, Yuya Seki, Shohei Watabe, Suguru Endo, Yuichiro Matsuzaki, Shiro
  Kawabata, Tetsuro Nikuni, and Hideaki Hakoshima.
\newblock Boltzmann machine learning with a variational quantum algorithm,
  2020.

\bibitem{Motta_2019}
Mario Motta, Chong Sun, Adrian T.~K. Tan, Matthew~J. O’Rourke, Erika Ye,
  Austin~J. Minnich, Fernando G. S.~L. Brandão, and Garnet Kin-Lic Chan.
\newblock Determining eigenstates and thermal states on a quantum computer
  using quantum imaginary time evolution.
\newblock {\em Nature Physics}, 16(2):205–210, Nov 2019.

\bibitem{McArdle_2019}
Sam McArdle, Tyson Jones, Suguru Endo, Ying Li, Simon~C. Benjamin, and Xiao
  Yuan.
\newblock Variational ansatz-based quantum simulation of imaginary time
  evolution.
\newblock {\em npj Quantum Information}, 5(1), Sep 2019.

\bibitem{2021yeterQITE}
Kübra Yeter-Aydeniz, Bryan~T. Gard, Jacek Jakowski, Swarnadeep Majumder,
  George~S. Barron, George Siopsis, Travis~S. Humble, and Raphael~C. Pooser.
\newblock Benchmarking quantum chemistry computations with variational,
  imaginary time evolution, and {K}rylov space solver algorithms.
\newblock {\em Advanced Quantum Technologies}, 4:2100012.

\bibitem{burges2002factoring}
Chris~J.C. Burges.
\newblock Factoring as optimization.
\newblock Technical Report MSR-TR-2002-83, January 2002.

\bibitem{devitt2005investigating}
Simon~J Devitt, Austin~G Fowler, and Lloyd~CL Hollenberg.
\newblock Investigating the practical implementation of shor's alagorithm.
\newblock In {\em Micro-and Nanotechnology: Materials, Processes, Packaging,
  and Systems II}, volume 5650, pages 483--494. International Society for
  Optics and Photonics, 2005.

\bibitem{PETZ_2011}
D.~PETZ and C.~GHINEA.
\newblock Introduction to quantum fisher information.
\newblock {\em Quantum Probability and Related Topics}, Jan 2011.

\bibitem{Qiskit}
H{\'e}ctor Abraham and AduOffei.
\newblock Qiskit: An open-source framework for quantum computing, 2019.

\bibitem{Yuan_2019}
Xiao Yuan, Suguru Endo, Qi~Zhao, Ying Li, and Simon~C. Benjamin.
\newblock Theory of variational quantum simulation.
\newblock {\em Quantum}, 3:191, Oct 2019.

\end{thebibliography}
